\documentclass[11pt,draftcls,onecolumn]{article}
\usepackage{multicol,caption}
\usepackage{multirow}

\usepackage[a4paper, total={7in, 9in}]{geometry}
\usepackage{graphicx}
\usepackage{changepage}
\usepackage{enumitem}
\usepackage{listings}
\usepackage{amsmath}
\usepackage{pdfpages}
\newenvironment{Figure}
{\par\medskip\noindent\minipage{\linewidth}}
{\endminipage\par\medskip}

\newcommand{\bc}{\begin{center}}
	\newcommand{\ec}{\end{center}}
\newcommand{\tab}[1][1cm]{\hspace*{#1}}

\begin{document}
	\title{Farmers' situation in agriculture markets and role of public interventions}
	\author{Vinay Reddy Venumuddala\\ PhD Student in Public Policy \\ Indian Institute of Management Bangalore.}
	\date{}
	\maketitle

	\begin{abstract}
		In our country, majority of agricultural workers (who may include farmers working within a cooperative framework, or those who work individually either as owners or tenants) are shown to be reaping the least amount of profits in the agriculture value chain when compared to the effort they put in. There is a good amount of literature which broadly substantiates this situation in our country.\\		
		\tab Main objective of this study is to have a broad understanding of the role played by public systems in this value chain, particularly in the segment that interacts with farmers. As a starting point, we first try to get a better understanding of how farmers are placed in a typical agriculture value chain. For this we take the help of recent seminal works on this topic that captured the situation of farmers' within certain types of value chains. Then, we isolate the segment which interacts with farmers and deep-dive into data to understand the role played by public interventions in determining farmers' income from agriculture. \\
		\tab NSSO 70th round on Situation Assessment Survey of farmers has data pertaining to the choices of farmers and the type of their interaction with different players in the value chain. Using this data we tried to get a econometric picture of the role played by government interventions and the extent to which they determine the incomes that a typical farming household derives out of agriculture.
	\end{abstract}

\section{Introduction}
\tab Agriculture is the most critical lifeline of India which forms the livelihood of more than 57.8\% of rural households, constituting close to 460 million persons in the country\textsuperscript{\cite{NSSO70SAS}}. Such a majority of the population is often found to be constantly pulled down by the gravity of deep rooted poverty and deprivation. For them to lead a life of dignity and satisfaction, it is essential to keep them out of those clutches which cause them to be pulled down under poverty. In this country, where the economy in Agriculture is found to greatly canvas over villages and their clusters, it is found that farmers, who form the core of this economy are stuck in a kind of weird network of forces which puts them on the top in terms of the contributions they make while not giving them a chance to reap profits on the same lines.\\
\tab  A study conducted by Kapur et.al\textsuperscript{\cite{kapur2014understanding}} found that farmers who are part of Red Gram Supply Chain in Karnataka contribute to more than 66\% of the costs of activities involved in the value chain, while reap less than 40\% of the total profits accrued in it. If one looks at the time that farmers spend in their contributing activities vis-a-vis other players, the intensity of their plight certainly rises. Often wholesale and retail traders, middle-men, commission agents, and people involved in processing and transport, who are part of this value chain are found only during the time of harvest but reap significant share in the profits accrued looking by all means, whether it be the time spent in earning these profits or the total amount they get by trading with a large number of farmers. In addition, farmers also bear the largest risks with little or no formal insurance or risk management options\textsuperscript{\cite{kapur2014understanding}}. In another study conducted by Sandip Mitra et.al\textsuperscript{\cite{mitra2017asymmetric}} on potato farmers in West Bengal, authors highlight the nature of asymmetric information in agriculture markets which contribute to a similar problem where farmers often end up in a relatively poor situation. Following paragraph quoted from their study gives a perspective of the problem at hand\\
\tab \textit{``Not only do farmers lack direct access to mandis; they are also unaware of the prices at which their potatoes are resold there. The gaps between these resale prices and farmgate prices are large: in the year of our study, farmgate prices were 44\% to 46\% of wholesale prices. Our calculations suggest that middlemen earned 50\% to 71\% of this gap. The pass-through from retail prices to farmgate prices was a statistically insignificant, negligible 2\%, while pass-through to wholesale prices was a much larger 81\%.''}\\
\tab Clearly this indicates that, within the network of various stakeholders who are part of agriculture value chain, it is farmers who often end up being the vulnerable lot. Although some big farmers and landlords may be exceptions in terms of having greater influence in the markets, their proportion in total farmers doesn't downplay the intensity of deprivation farmers' as a group face in this country. According to NSSO Situation Assessment Survey report\textsuperscript{\cite{NSSO70SAS}} of 2013,\\
\tab \textit{``About 69 percent of the agricultural households possessed land less than 1 hectare during the agricultural year July 2012- June 2013. Only 0.4 percent of the agricultural households possessed land 10 hectares or more.''}\\
\tab In the following sections, we first try to understand the situation of farmers in the context of agriculture value chain from few recent studies. Then we carefully try to look at the segment within this value chain where farmers usually make their choices which determine the extent of profits they may receive out of agriculture. The key reflections of farmers' choices or their determinants that we investigate from data are \textit{a. Information inputs (Awareness of MSP and source of technical advice), b. Choices of Input side procurement (Whether procured from Govt/Non-Govt agencies), and c. Choice for Sale of produce(Whether sale to Govt/Non-Govt agencies)}. We discuss the role of government interventions in this context by an econometric analysis of how farmers' choices or their determinants listed previously, contribute to their agriculture incomes. 

\section{Agriculture markets and role of public interventions}
Public interventions into agriculture find their place mainly in strengthening forward and backward linkages that connect farmers to the input and produce markets. On the input side public institutions play a key role in helping farmers with their choice of seeds, fertilizers, technology etc.., in farming. On the output/produce side they have two major roles. Declaring and enabling Minimum Support Price (MSP) for certain notified agricultural commodities is one role where the public institutions play a major role. In procurement of the said kinds of agriculture produce through government channels, these institutions act as enabling systems for realizing MSP. The second and also a major role that is taken up by the state governments\footnote{Agriculture is a state subject}, is to regulate agricultural commodity markets through state specific Agriculture Produce Marketing (APM) Acts. Within the purview of these acts, corresponding public systems (Here, Agriculture Produce Marketing Committees) manage Mandi operations and govern the way trade is carried out within them. \\
\tab In this study, we mainly focus on the role played by government in the inputs' side of agriculture and some aspects in the output side mainly in its role as a procurement agent. Firstly, on the inputs' side, in terms of information dissemination in agriculture, public systems interact with farmers in determining their choice of technology, choice of crop and other practices. They also facilitate programs which may determine farmers' awareness of MSP. Secondly, government interventions may also determine the choice of procurement of inputs like seeds and fertilizers. Lastly, in the output side, public systems play the role of procurement agent alongside non-government players like local traders, middle-men and other agents which procure farm produce.\\
\tab In the subsequent sections, using NSSO data (Situation Assessment Survey data of 2013 (NSSO-SAS)\textsuperscript{\cite{NSSO70SAS}}), we broadly investigate into the role played by public systems or government interventions particularly on the above mentioned dimensions in contributing to farmers' agricultural incomes. Before proceeding to econometric analysis on the available data, in the following section, we first look at the placement of farmers' in a typical agriculture value chain from recent seminal works, which may help us further to relate the results and discuss in context.

\section{Farmers in agriculture value chains - a review of literature}
In their study on Understanding price variation in agriculture commodities in India, Chatterjee et.al\textsuperscript{\cite{chatterjee2016understanding}} have given a lucid narrative of the history of agriculture markets in India. Following paragraph summarizes an existing governance mechanism in place, in terms of operation of Mandis, which play a dominant role in agriculture markets. \\
\tab ``\textit{State specific APM Acts provide for the formation of Agriculture Produce Marketing Committees (APMC) which are tasked to operate agriculture markets. Prices are discovered through what in principle is an open auction. Critically, once an area was declared a market area and falls under the jurisdiction of a Market Committee, no person or agency was allowed freely to carry on wholesale marketing activities elsewhere. Not only did the government issue licenses to trade in these markets but also the licenses were state and mandi specific. First sale in the notified agricultural commodities produced in the region such as cereals, pulses, edible oilseed, fruits and vegetables and even chicken, goat, sheep, sugar, fish etc., can be conducted only under the aegis of the APMC, through its licensed commission agents, and subject to payment of various taxes and fee.}\textsuperscript{\cite{chatterjee2016understanding}}''.\\ \tab However in the implementation of these acts as regulatory agents, many governments failed to ensure that farmers were offered fair prices in a transparent manner. These failures were mainly attributed to severe governance challenges in mandis which caused to bring down these markets as fragmented, inefficient and lacking transparency in their operations\textsuperscript{\cite{chatterjee2016understanding}}.\\
\tab In another study, Kapur et.al\textsuperscript{\cite{kapur2014understanding}}, along with IIM Bangalore, have undertaken research to analyze Red Gram supply chain in Karnataka. Following picture taken from their Working paper on Understanding Mandis\textsuperscript{\cite{kapur2014understanding}}, show the various players involved, and their interactions within red gram supply chain in Karnataka. Their study revealed that, on average, the farmers' profit comes out to be approximately 39\% of the profit across the supply chain, in contrast to their contribution which is found to be over 66\% of the total supply chain costs.\\

\begin{Figure}
	\captionsetup{font=scriptsize}
	\begin{center}
		\includegraphics[width=5.0in]{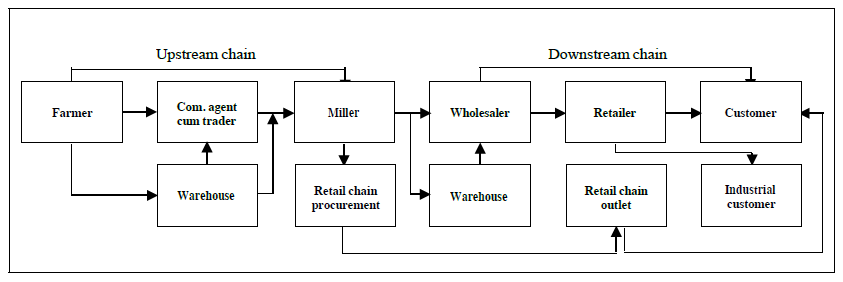}
		\captionof{figure}{Red Gram Supply Chain} 
		\label{fig:RGSC}
	\end{center}	
\end{Figure}

In another important study by Sandip Mitra et.al\textsuperscript{\cite{mitra2017asymmetric}}, an extensive analysis through thorough field research has been carried out to understand the nature of trading relationships between middlemen and farmers. Farmers in the context of their study\footnote{Potato Farmers} sell most of their produce to village middlemen, who aggregates it and resells at wholesale markets to bulk buyers from distant cities or neighbouring states. Having no direct access to Mandis, or information about prices there, these farmers conduct their trade interacting with either village middlemen or local traders from nearby villages. Farmers here have options of either selling to village middlemen, or these traders in local haat\footnote{A  market place organized mostly at cluster (a group of villages) level}, or to send their produce to government provided storage facility. This lack of direct access to wholesale markets is distinctive to West Bengal potato supply chain, and the situation in other states might be different\textsuperscript{\cite{mitra2017asymmetric}}. However, the situation in terms of interacting with middle-men and traders might be similar across other states as well. \\
\tab In all these studies one common phenomenon that has been observed, is that farmers (particularly small and marginal) are at a considerable disadvantage in terms of the profits they get in the agriculture value chains in comparison to the effort they put in. With a major contribution in the value chain, producers reaping a compromisingly low profits is a therefore worry-some scenario in the country.\\
\tab Above studies also discuss in detail on the nature of interaction of farmers' with different players in the agriculture value chain like, government, local traders, middle-men and so on. A glimpse of which has been discussed in the previous paragraphs. However, on the input side, in terms of farmers' procuring seeds, fertilizers and other requirements, a similar nature of interaction persists with these players. Many times farmers' procure inputs and also information from the very same players to whom they sell their produce. This may be due to dependence with the latter in terms of financial support or other reasons. 

\section{Farmers' choices and role of public interventions - contribution to agricultural income}
\tab Above discussion gives a perspective from prior studies on the possible factors that may play role in determining farmers' income from agriculture. Many such studies help in providing a rather worm's-eye perspective of the farmers' situation, but may not truly give a macro picture. In order to understand the farmers' situation and the extent to which role of public interventions mattered across country, we need to quantitatively get such a picture from available country-wide datasets. NSSO Situation Assessment Survey (SAS) of agricultural households, the recent one conducted as a part of 70th round under 33rd Schedule, has pertinent information on agricultural household characteristics which we try to use in this study. Using this data, we intend to conduct econometric analysis in trying to answer the following questions.\vspace{0.2cm} \\ 
\textit{\textbf{Q1}. To what extent does the role of government in providing access to information (about Minimum Support Price and technical advice regarding farming practices) help in contributing to farmers' agriculture income?\\
\textbf{Q2}. To what extent, does the role of public systems in provision of agricultural inputs (mainly seeds and fertilizers), contribute to the agriculture incomes of farming households? and \\
\textbf{Q3}. What is the difference in agriculture incomes possibly caused due to the sale of output to government vs non-government agencies?\\}
\section{Econometric Analysis}
In this section, we use econometric techniques to understand the role of public interventions, based on NSSO SAS data. For the purpose of simplicity, we illustrate the analysis pertaining to each question in the following subsections, and present the results at the end. In all our models total income from agriculture excluding animal husbandry (or Total Income solely from Cultivation of Crops ($TIC_i$)) is used as dependent variable in our analysis. This can be computed by the difference of total crop sale value ($TSV_i$) and total crop expenses ($TCE_i$) available for each household in the dataset. \footnote{For the purpose of our study, we chose to work on Situation Assessment Survey data pertaining to Visit-1 alone, i.e during the agriculture season from July to December 2012. However, since the sample for Visit-2 (Jan to June 2012) is almost the same (except little less in number), we assume that the results discussed based on Visit-1 may be more or less valid for the other agriculture season as well.}\\
\tab $TIC_i = TSV_i - TCE_i$\\
We truncate this variable at zero, so that households having net income less than zero are treated to have zero income. Essentially we have truncated the dependent variable here. Proportion of households having net income less than zero is close to 7\% in the data which is very small compared to the total number of households, therefore for the purpose of this study we ignore the effect of truncation. After accounting for incomes less than zero, we finally use $log(1+TIC_i)$ as our dependent variable in the regression models discussed further, because typically having a logarithm of incomes is beneficial to account for a wide variation in income levels. 
\subsection{Role of public interventions - Information inputs}
To answer \textit{\textbf{Q1}}, we first create the relevant variables signifying awareness of farmers about MSP (\textit{MSP\_Aware}) and source of technical information about agriculture (Whether from Govt/Non-Govt agencies, \textit{Tech\_Govt}). We then address two different sub-questions here, one is by addressing the impact of awareness of MSP on agriculture income and second is by addressing the impact of source of technical advice on agriculture income of farmers. We use propensity score matching technique to answer each of these sub-questions and the subsequent questions discussed in later sections.\footnote{ Essentially this technique helps in matching the background characteristics of treatment group and control group, for whichever treatment we plan to analyze. In this case, we consider treatment to be a binary variable which signifies either a. Awareness of farmers about MSP (\textit{MSP\_Aware}, which takes values 0,1) or b. Source of technical advice (\textit{Tech\_Govt}, which takes values 0,1) respectively for the two sub-questions we plan to address.}
\subsubsection{Awareness of MSP}
We use \textit{MSP\_Aware} as the grouping variable, and \textit{DCV\_b5a\_q13, CropCode, Tech\_Govt, InAgent\_Govt, State, b3\_q1,2, and 3} as our matching variables for analyzing the impact of awareness of MSP on agriculture incomes. Following are the list of constructed variables and their description, in the models shown in this section and further in our study. \vspace{0.3cm}
\begin{table}[!htbp] \centering 
			\begin{tabular}{|p{0.5in}|p{2in}|p{4in}|} 
				\hline
				\textbf{S.No}&	\textbf{Variable Name}&	\textbf{Description}\\ \hline
				1&\textit{DCV\_b5a\_q13}&Total land used for agriculture\\ \hline
				2&\textit{\textbf{CropCode}}&Choice of Major Crop during the season from July-December 2012 (Vector of Crop Dummies) \\ \hline
				3&\textit{b4q7}&Indicates whether any member of the household attended formal training in agriculture \\ \hline
				4&\textit{Tech\_Govt}&Indicates whether source of technical advice is from Govt/Non-Govt agencies \\ \hline
				5&\textit{InAgent\_Govt}&Indicates whether seeds or fertilizer inputs are procured from Govt/Non-Govt agencies \\ \hline
				6&\textit{pAgent\_Govt}&Indicates whether produce is sold to Govt/Non-Govt agencies \\ \hline
				7&\textit{MSP\_Aware}&Indicates whether a farming household is aware of MSP or not\\ \hline
				8&\textit{lnTIC}&$log(1+TIC_i)$ as mentioned in the start of this section\\ \hline
				9&\textit{\textbf{State}}&Variable accounting for state fixed effects (Vector of State Dummies)\\ \hline
				10&\textit{b3\_q1}&Household Size\\ \hline
				11&\textit{b3\_q2}&Religion Code\\ \hline
				12&\textit{b3\_q3}&Social Group\\ \hline
			\end{tabular}
	\captionsetup{font=scriptsize,singlelinecheck=off,justification=centering}
	\captionof{table}{List of constructed variables and their description for the purpose of study}
	\label{tab:VarCons}
\end{table}

Using propensity score matching, we use the matched dataset\footnote{Treatment and Control Group background characteristics (explanatory variables excluding the grouping variable in the corresponding regression) are matched, and resultant dataset is used for our regression model in each case} to run the following regression to capture the effect of awareness of MSP on the agriculture incomes. \textit{(Note: We did not use variables \textit{b4\_q7}, \textit{Tech\_Govt} and \textit{PAgent\_Govt} as these variables most likely are collinear to our treatment variable and hence matching on the basis of these might give unreliable estimates)} \\
\tab \textit{lnTIC} = $\beta_0$+$\beta_1$ \textit{MSP\_Aware} + $\epsilon$ -- (1a),\\ Which essentially captures the effect of awareness of MSP in farmers' agriculture incomes. 
\subsubsection{Source of technical advice}
Similar to above model, we use \textit{Tech\_Govt} as the grouping variable here, and other matching variables are the same as that of the previous model. We include an additional matching variable \textit{PAgent\_Govt} here.\footnote{Unlike in Awareness of MSP, \textit{PAgent\_Govt} need not be collinear with the variable \textit{Tech\_Govt}. Because those who are aware of MSP alone will be able to sell to Govt agencies, while those who receive technical advice need not receive them from Govt agencies alone.} The regression model after balancing the matching variables is \\
\tab \textit{lnTIC} = $\beta_0$+$\beta_1$ \textit{MSP\_Aware} + $\epsilon$ -- (1b),\\ Which essentially captures the effect of source of technical advice from government in farmers' agriculture incomes (As we controlled for other background characteristics using propensity score matching).

\subsection{Role of public interventions - Provision of agricultural inputs}
In order to understand the role of public interventions in terms of provision of agricultural inputs (for addressing \textbf{\textit{Q2}}), we construct a variable \textit{InAgent\_Govt}, whose description is given in Table-1. We use this as grouping variable or treatment variable for our model, and \textit{DCV\_b5a\_q13, CropCode, b4\_q7, Tech\_Govt, PAgent\_Govt, State, b3\_q1,2, and 3} as our matching variables. As we can see, we are essentially controlling for background characteristics like choice of crop, land usage, source of technical advice, choice of output sale, state and HH fixed effects. Following regression model gives us the effect of government facilitated inputs on agriculture incomes.\\
\tab \textit{lnTIC} = $\beta_0$+$\beta_1$ \textit{InAgent\_Govt} + $\epsilon$ -- (2)

\subsection{Role of public interventions - Provision of agricultural inputs}
In order to understand the role of public interventions in terms of provision of procurement support (for addressing \textbf{\textit{Q3}}), we construct a variable \textit{PAgent\_Govt}, whose description is given in Table-1. We use this as grouping variable or treatment variable for our model, and \textit{DCV\_b5a\_q13, CropCode, b4\_q7, Tech\_Govt, 	InAgent\_Govt, State, b3\_q1,2, and 3} as our matching variables. As we can see, we are essentially controlling for background characteristics like choice of crop, land usage, source of technical advice, Choice of input suppliers, state and HH fixed effects. Following regression model gives us the effect of government facilitated procurement support on agriculture incomes.\\
\tab \textit{lnTIC} = $\beta_0$+$\beta_1$ \textit{PAgent\_Govt} + $\epsilon$ -- (3)

\subsection{Results}
Following tables-\ref{tab: Result1ab} and \ref{tab: Result23} gives the regression results\footnote{We use stargazer package\textsuperscript{\cite{stargazer}} in R for printing the regression results in this format} of the four models explained above. Figures depicting corresponding propensity score distributions before and after matching is shown in the appendix section.
\begin{table}[!htbp] \centering 
	\caption{Regression Results - (1a) and (1b)} 
	\label{tab: Result1ab} 
	\begin{tabular}{@{\extracolsep{5pt}}lcc} 
		\\[-1.8ex]\hline 
		\hline \\[-1.8ex] 
		& \multicolumn{2}{c}{\textit{Dependent variable:}} \\ 
		\cline{2-3} 
		\\[-1.8ex] & \multicolumn{2}{c}{lnTIC} \\ 
		\\[-1.8ex] & Result1a & Result1b\\ 
		\hline \\[-1.8ex] 
		MSP\_Aware & 0.471$^{***}$ &  \\ 
		& (0.060) &  \\ 
		Tech\_Govt &  & 0.248$^{***}$ \\ 
		&  & (0.072) \\ 
		Constant & 9.379$^{***}$ & 9.489$^{***}$ \\ 
		& (0.042) & (0.051) \\ 
		\hline \\[-1.8ex] 
		Observations & 9,106 & 7,096 \\ 
		R$^{2}$ & 0.007 & 0.002 \\ 
		Adjusted R$^{2}$ & 0.007 & 0.002 \\ 
		Residual Std. Error & 2.853 (df = 9104) & 3.025 (df = 7094) \\ 
		F Statistic & 62.125$^{***}$ (df = 1; 9104) & 11.967$^{***}$ (df = 1; 7094) \\ 
		\hline 
		\hline \\[-1.8ex] 
		\textit{Note:}  & \multicolumn{2}{r}{$^{*}$p$<$0.1; $^{**}$p$<$0.05; $^{***}$p$<$0.01} \\ 
	\end{tabular} 
\end{table}

\begin{table}[!htbp] \centering 
	\caption{Regression Results - (2) and (3)} 
	\label{tab: Result23} 
	\begin{tabular}{@{\extracolsep{5pt}}lcc} 
		\\[-1.8ex]\hline 
		\hline \\[-1.8ex] 
		& \multicolumn{2}{c}{\textit{Dependent variable:}} \\ 
		\cline{2-3} 
		\\[-1.8ex] & \multicolumn{2}{c}{lnTIC} \\ 
		\\[-1.8ex] & Result2 & Result3\\ 
		\hline \\[-1.8ex] 
		InAgent\_Govt & $-$0.149$^{*}$ &  \\ 
		& (0.083) &  \\ 
		PAgent\_Govt &  & 0.239$^{*}$ \\ 
		&  & (0.122) \\ 
		Constant & 9.389$^{***}$ & 10.222$^{***}$ \\ 
		& (0.059) & (0.086) \\ 
		\hline \\[-1.8ex] 
		Observations & 5,710 & 1,540 \\ 
		R$^{2}$ & 0.001 & 0.002 \\ 
		Adjusted R$^{2}$ & 0.0004 & 0.002 \\ 
		Residual Std. Error & 3.129 (df = 5708) & 2.397 (df = 1538) \\ 
		F Statistic & 3.238$^{*}$ (df = 1; 5708) & 3.832$^{*}$ (df = 1; 1538) \\ 
		\hline 
		\hline \\[-1.8ex] 
		\textit{Note:}  & \multicolumn{2}{r}{$^{*}$p$<$0.1; $^{**}$p$<$0.05; $^{***}$p$<$0.01} \\ 
	\end{tabular} 
\end{table} 
From these results, we see that the role of public systems in terms of awareness and capacity building are significant in improving agriculture income of farmers. While their role in terms of acting as agents in facilitating input supply or output procurement is only slightly significant. Inferences that come up from the latter results are that, public systems acting on the input supply side are somehow negatively affecting the agriculture incomes, while their role as output procurement agents is positive.

\section{Discussion and Future Work}
Although above results might not reveal a complete picture of that segment in agriculture value chain, in which farmers' play critical role, nevertheless it helps us get a macro picture of how government interventions play out in this segment. A major take away seems to be that, interventions like enhancing the farmers' awareness in terms of availing MSP, and enhancing capabilities of farmers through pertinent technical advice, contribute positively and significantly to the agriculture income of farming households. \\
\tab The second and another important result \footnote{Discussed in Results section \textit{- Public systems acting on the input supply side are somehow negatively affecting the agriculture incomes, while their role as output procurement agents is positive.}}, seems to match with that of an existing study by Chatterjee et.al\textsuperscript{\cite{chatterjee2016understanding}}, where in the context of paddy procurement, it is found that government intervention improves trade in favour of farmers. It seems like the slightly significant result showing a positive effect of the role played by public procurement in improving agricultural incomes is broadly true when we look at an overall picture of agriculture in the country. However, we couldn't find relevant studies which might look into input side government interventions and their manifestations at all India level, to compare our results with. In any case, at first sight, the results seem to convey a negative effect on farmers' agriculture incomes due to government interventions in input side markets. A further deep-dive analysis is essential here to find out the possible reasons behind this phenomenon. \\
\tab In this study we have looked mainly into the segment of agriculture value chain where farmers are involved. However, a standalone picture of this sort might not help in analyzing the overall importance of public systems or government interventions in the complete value chain. Role played by government in other segments of the value chain also need to be similarly analyzed and integrated, to essentially understand the gaps in public systems management in the context of agriculture value chains. Thorough research of this sort may help policy makers to get a decent enough direction to look for improvement of public systems in agriculture. \\
\tab Also in this study, we limited ourselves to a binary view in terms of role played by Government or non-government agencies in contributing to agriculture incomes. However, one can go further and see what all players within the government and non-government agencies are actually contributing efficaciously to the overall value chain. This might be an interesting and important study one can look at. In conclusion, through this study we recommend researchers to take a public management perspective of looking at policy networks within agriculture value chains, to better understand the role played by public systems at large. Thorough research on these lines can build a strong foundation in improving the overall effectiveness of these systems over a longer horizon.

\section{Appendix}
\subsection{Distribution of propensity scores before and after matching}
\begin{Figure}
	\captionsetup{font=scriptsize}
	\begin{center}
		\includegraphics[width=4.0in]{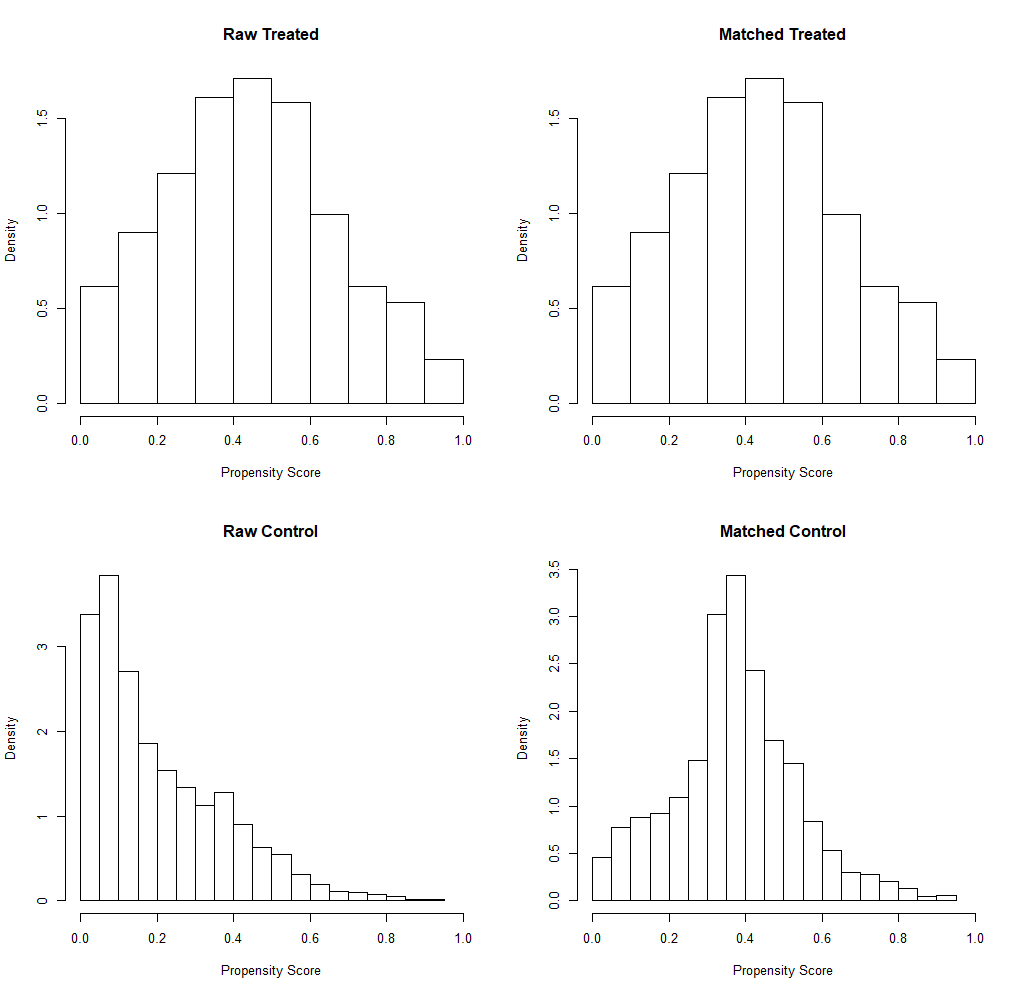}
		\captionof{figure}{Treatment: Awareness of MSP} 
		\label{fig:result1a}
	\end{center}	
\end{Figure}

\begin{Figure}
	\captionsetup{font=scriptsize}
	\begin{center}
		\includegraphics[width=4.0in]{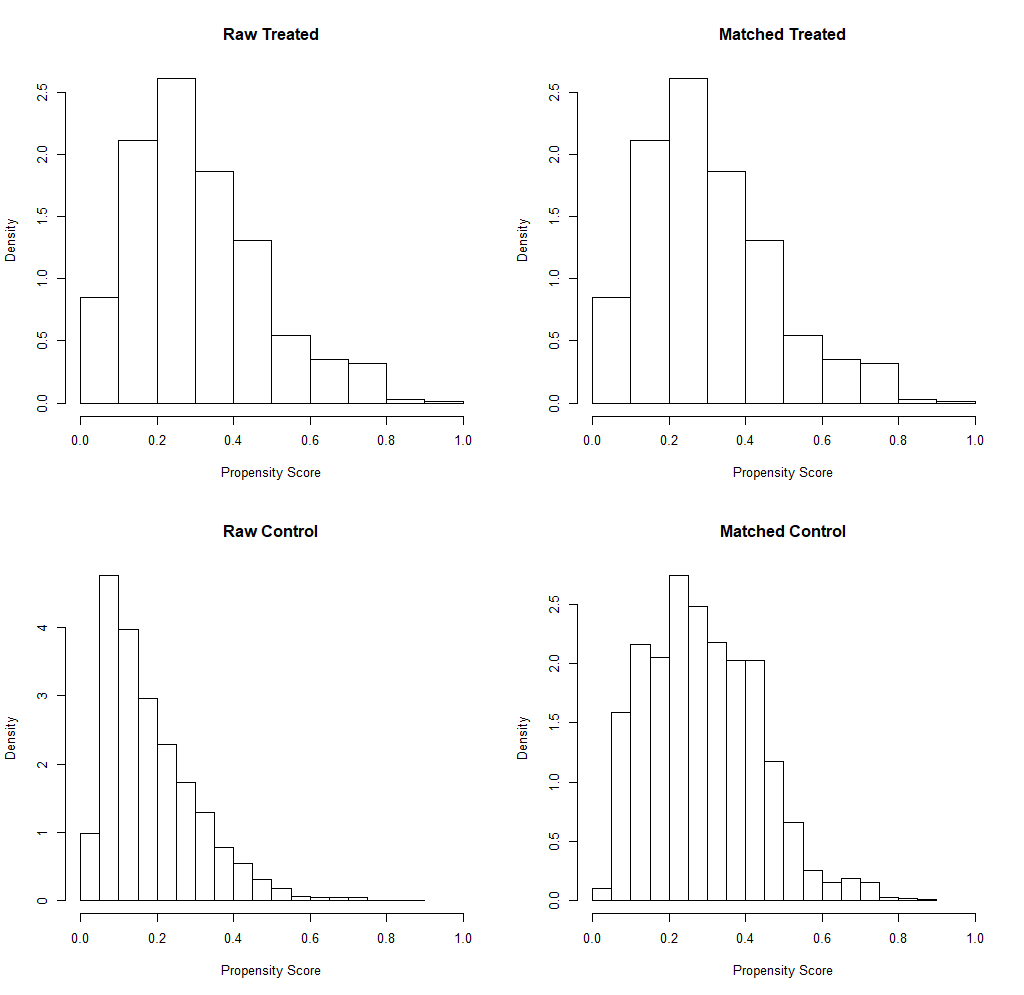}
		\captionof{figure}{Treatment: Source of Technical Advice (Govt/Non-Govt)} 
		\label{fig:result1b}
	\end{center}	
\end{Figure}

\begin{Figure}
	\captionsetup{font=scriptsize}
	\begin{center}
		\includegraphics[width=4.0in]{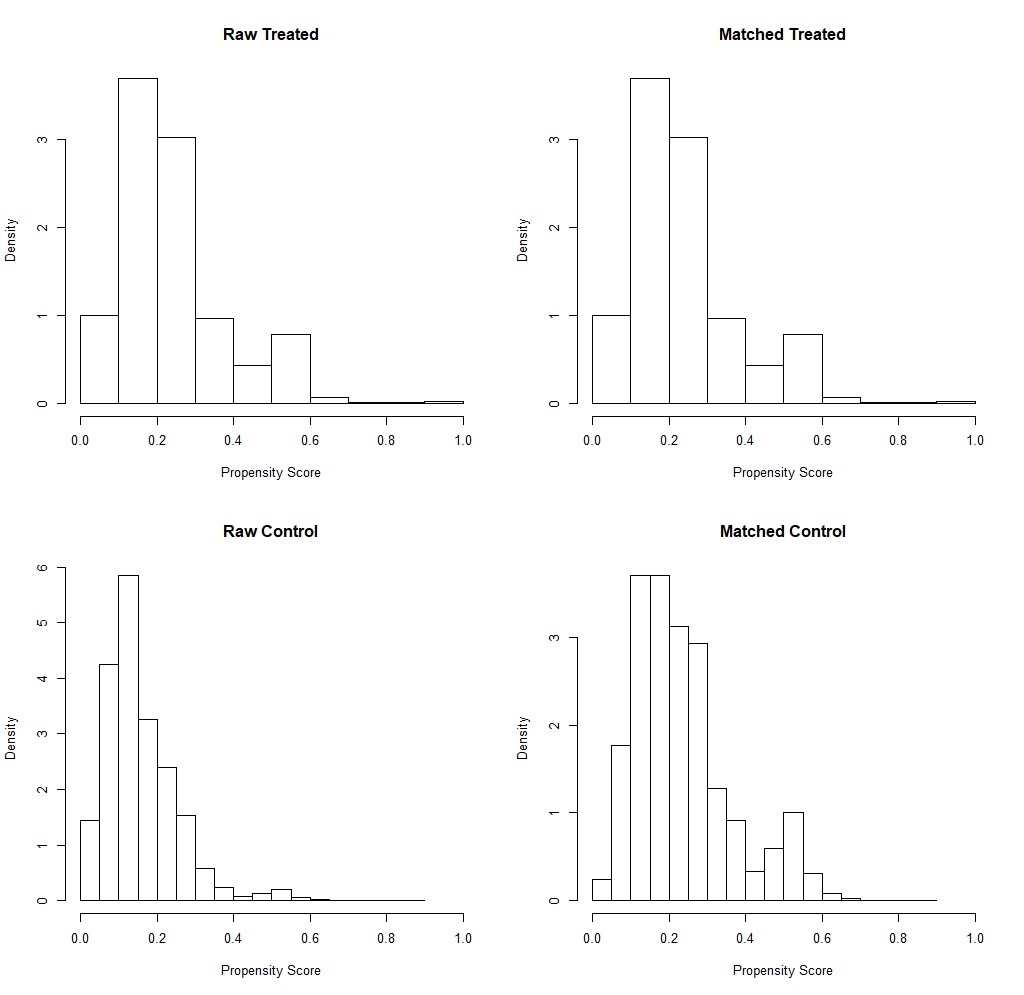}
		\captionof{figure}{Treatment: Inputs from (Govt/Non-Govt)} 
		\label{fig:result2}
	\end{center}	
\end{Figure}

\begin{Figure}
	\captionsetup{font=scriptsize}
	\begin{center}
		\includegraphics[width=4.0in]{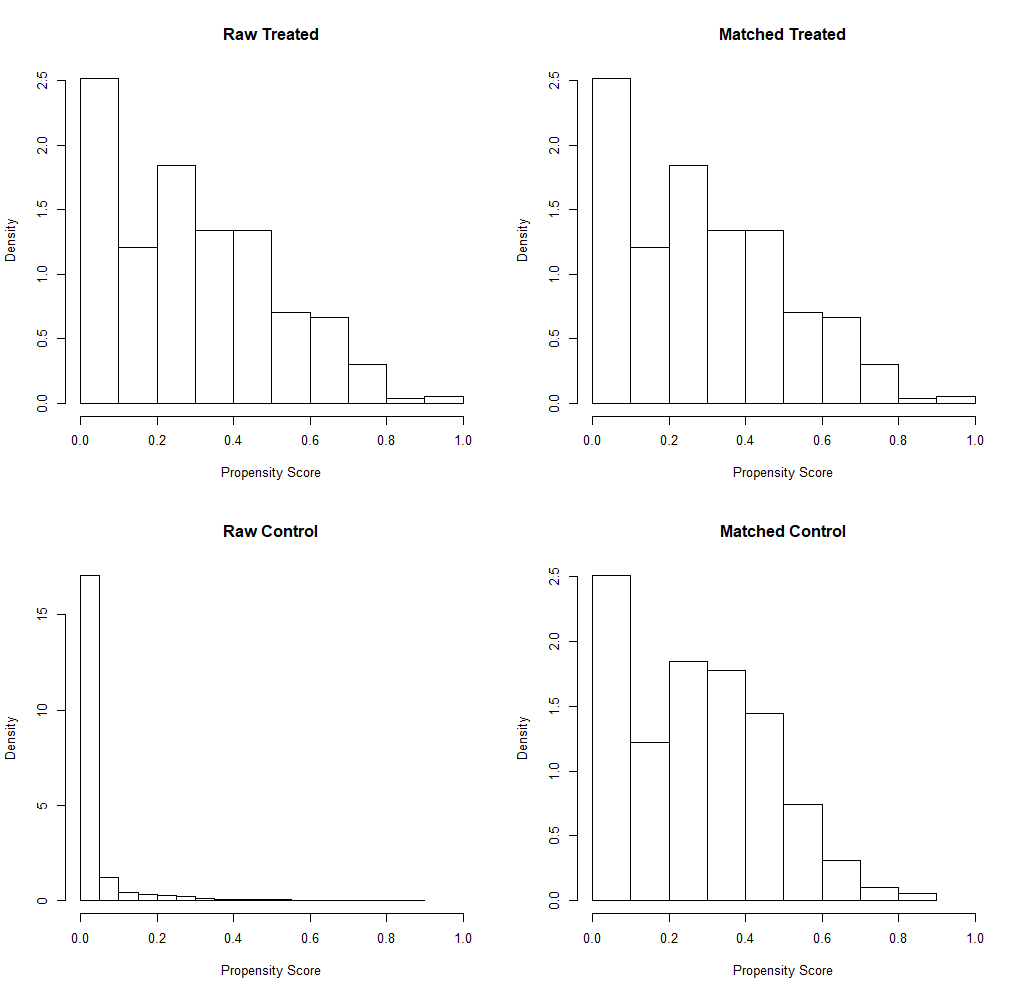}
		\captionof{figure}{Treatment: Output sale to (Govt/Non-Govt)} 
		\label{fig:result3}
	\end{center}	
\end{Figure}

	\bibliographystyle{Generic}
	\bibliography{reflibrary}

	
\end{document}